\documentclass[letterpaper,10pt]{article} 

\usepackage{osameet3} 

\newcommand\authormark[1]{\textsuperscript{#1}}

\usepackage{amsmath,amssymb}
\usepackage[colorlinks=true,bookmarks=false,citecolor=blue,urlcolor=blue]{hyperref} 
\usepackage{float} 
\usepackage{siunitx} 

\begin{document}

\title{Limited-memory BFGS Optimisation of Phase-Only Computer-Generated Hologram for Fraunhofer Diffraction}

\author{Jinze Sha*,\authormark{1} Andrew Kadis,\authormark{1} Fan Yang,\authormark{1} Timothy D. Wilkinson\authormark{1}}

\address{\authormark{1}Electrical Engineering Division, Department of Engineering, University of Cambridge, 9 JJ Thomson Avenue, Cambridge CB3 0FA, United Kingdom}

\email{\authormark{*}js2294@cam.ac.uk} 

\begin{abstract}
We implement a novel limited-memory Broyden–Fletcher–Goldfarb–Shanno (L-BFGS) optimisation algorithm with cross entropy (CE) loss function, to produce phase-only computer-generated hologram (CGH) for holographic displays, with validation on a binary-phase modulation holographic projector.
\end{abstract}

\section{Introduction}
This paper demonstrates a novel L-BFGS algorithm with CE loss function to generate phase-only CGH for holographic displays for far-field (using Fraunhofer diffraction), along with Adam algorithm and mean squared error (MSE) loss function for comparison, and the hologram generated is validated on a binary-phase modulation holographic projector. The inspiration of using L-BFGS optimisation algorithm on CGH arose from the previous work on Recurrent Neural Network Language Models (RNNLM) \cite{8461550}, where the combination of L-BFGS algorithm with CE loss function demonstrated outstanding performance on RNNLM. Most previous work on optimisation of CGH are for Fresnel diffraction (in the near-field) instead of Fraunhofer diffraction  (in the far-field), despite Ref. \cite{app10124283}, where a non-stochastic gradient descent optimisation of CGH for Fraunhofer diffraction is demonstrated.

\section{Theory}
Fraunhofer diffraction is a form of diffraction in which the distance between the light source and the receiving screen are in effect at infinite, so that the wave fronts can be treated as planar rather than spherical \cite{Fraunhoferdiffraction}. 

The optimisation algorithms used in this paper are Adam and L-BFGS. Adam algorithm is a first-order gradient-based optimisation of stochastic objective functions, based on adaptive estimates of lower-order moments \cite{Adam:14}. L-BFGS algorithm is a quasi-Newtonian method which determines the gradient with curvature information \cite{Liu1989OnTL}, from the gradient history. In this paper, a gradient history of size 20 is used throughout the experiments.

The loss functions used in this paper are MSE and CE. MSE is calculated as shown in Eq. \ref{eq:MSE}. MSE is a traditional metric averaging the squared error between target values $X_i$ and observed valued $Y_i$, where $n$ is the size of $X$ and $Y$. CE is calculated as shown in Eq. \ref{eq:CE}. Unlike MSE, CE averages the products of $X_i\log(Y_i)$. CE is often used in classification problems. An example is language modelling \cite{8461550}, where the cross-entropy is used to assess the error of the language model. CE is computed using the true distribution of words ($X$), and the predicted distribution of words by the model ($Y$). In this paper, as the loss functions are applied to two-dimensional (2D) images, the 2D images need to be flattened into one-dimensional (1D) array before calculation.

\vspace{-2mm}
\begin{minipage}{0.45\textwidth}
    \begin{eqnarray}
    MSE(X,Y)=\frac{1}{n}\sum_{i=1}^{n}(X_i-Y_i)^2
    \label{eq:MSE}
    \end{eqnarray}
\end{minipage}\hfill
\begin{minipage}{0.45\textwidth}
    \begin{eqnarray}
    CE(X,Y)=-\frac{1}{n}\sum_{i=1}^{n}X_{i}\log(Y_{i})
    \label{eq:CE}
    \end{eqnarray}
\end{minipage}
\vspace{-2mm}
\begin{figure}[H]
  \centering
  \includegraphics[width=7cm]{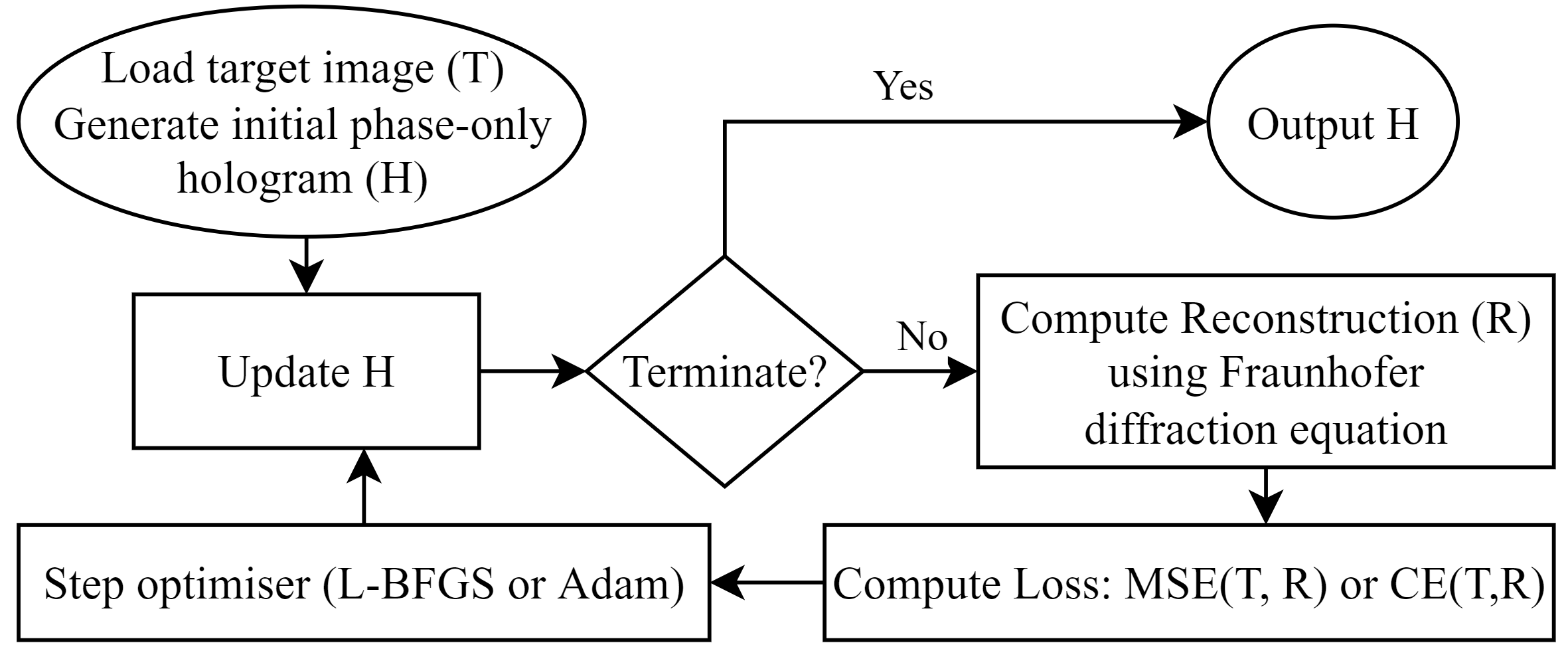}
  \vspace{-3mm}
  \caption{Flowchart for optimisation of Phase-Only CGH}\label{fig:OptiCGH-Flowchart}
\end{figure}
\vspace{-5mm}

The flowchart of the optimisation process is shown in Fig. \ref{fig:OptiCGH-Flowchart}. In order to set up optimisation for CGH, the objective function to minimise needs to be the loss function between the target image ($T$) and reconstructed image ($R$) using Fraunhofer diffraction. And the optimising argument is the phase-only hologram ($H$).

\section{Experiment}

\vspace{-4mm}
\begin{figure}[H]
  \centering
  \includegraphics[width=0.8\textwidth]{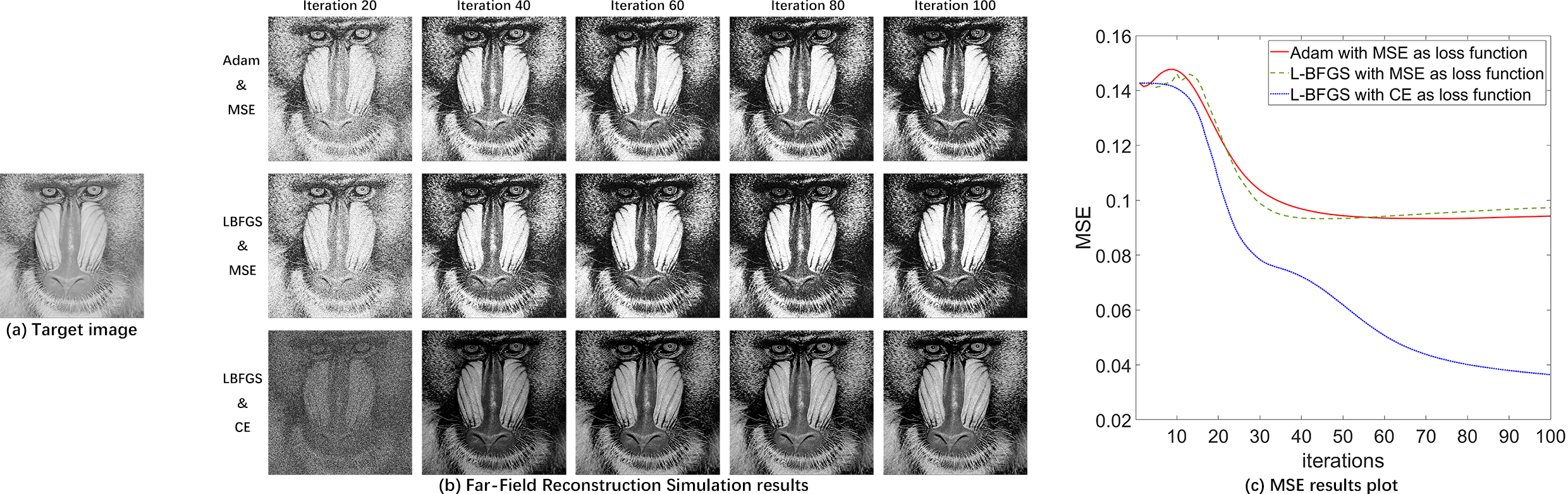}
  \vspace{-3mm}
  \caption{Simulation results and MSE plot}\label{fig:Simulation-results}
\end{figure}
\vspace{-5mm}

The theory was then implemented. Aiming for a fairer comparison, the learning rate parameters were all set to 0.1 and the number of iterations were all set to 100, and all iterations started from the same randomly generated initial phase-only hologram. The target image used is the mandrill as shown in Fig. \ref{fig:Simulation-results} (a).

All combinations of the two optimisation algorithms (L-BFGS and Adam) and the two loss functions (MSE and CE) were tried, among which the Adam optimisation with CE as loss function did not converge, while the other three combinations converged successfully. For each of the three combinations, the reconstructions ($R$) are saved at each iteration and every 20 of them are shown in Fig. \ref{fig:Simulation-results} (b). The MSE of all $R$'s are plotted in Fig. \ref{fig:Simulation-results} (c), where it can be seen that, L-BFGS with CE as loss function not only converges faster (in number of iterations), but also results in much lower MSE (around 0.035) than the others (around 0.095), demonstrating a significant advantage.

\vspace{-2mm}
\begin{figure}[H]
  \centering
  \includegraphics[width=7cm]{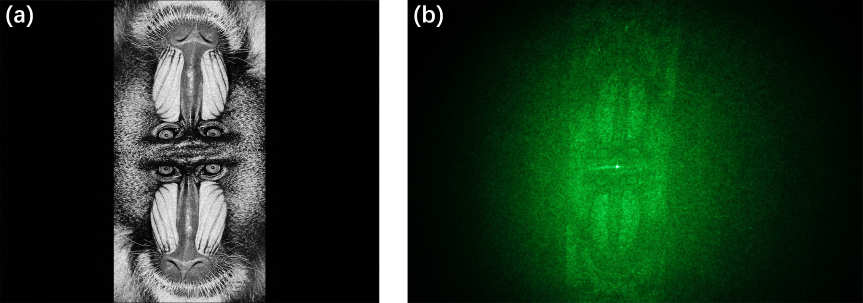}
  \vspace{-3mm}
  \caption{Experiment Results. (a) Simulation with 180$^{\circ}$ rotational symmetry, (b) Replay field captured.}\label{fig:Experiment}
\end{figure}
\vspace{-5mm}

The holographic projector in the experiment is equipped with a binary-phase ferroelectric liquid crystal (FLC) spatial light modulator (SLM) with 1440Hz refresh rate, 1280x1024 resolution, and \SI{13.6}{\micro\metre} pixel pitch, illuminated by a diode-pumped solid-state (DPSS) laser source of $532nm$ wavelength and $50mW$ power \cite{Freeman:09}. The nature of the binary-phase modulation restricts that there is always a 180$^{\circ}$ rotational symmetry in the replay field, so only half the plane is useful. Hence the target image is designed to be rotational symmetrical, and after 100 iterations of L-BFGS optimisation with CE as loss function, the reconstruction simulated is shown in Fig. \ref{fig:Experiment} (a). After uploading the according hologram to the SLM, the replay field was then captured as shown in Fig. \ref{fig:Experiment} (b), using a Canon 550D camera with an EFS 18-55mm lens. The image of the replay field validates the CGH, despite some noise presenting, partly caused by the binary quantisation of the hologram.

\vspace{-1mm}
\section{Conclusion and future work}
\vspace{-1mm}
The simulation compared the effectiveness of L-BFGS and Adam optimisation algorithms with CE and MSE as loss function to generate phase-only CGH for Fraunhofer diffraction. L-BFGS optimisation with CE loss function was found to be the best, and the hologram produced was validated experimentally. Further work is required to quantify the quality of the replay field, to compare the running time of the algorithms and to get better result by adjusting learning rate and number of iterations. L-BFGS algorithm can also be extended for Fresnel diffraction.

\vspace{-1mm}

\bibliographystyle{osajnl}
\bibliography{OptimisationCGH}

\end{document}